\documentclass[pdflatex,sn-mathphys-num]{sn-jnl}

\usepackage{graphicx}%
\usepackage{multirow}%
\usepackage{amsmath,amssymb,amsfonts}%
\usepackage{amsthm}%
\usepackage{mathrsfs}%
\usepackage[title]{appendix}%
\usepackage{xcolor}%
\usepackage{textcomp}%
\usepackage{manyfoot}%
\usepackage{booktabs}%
\usepackage{algorithm}%
\usepackage{algorithmicx}%
\usepackage{algpseudocode}%
\usepackage{listings}%

\theoremstyle{thmstyleone}%
\theoremstyle{thmstyletwo}%

\theoremstyle{thmstylethree}%

\begin{document}

\title[Article Title]{Computation in flight through precision-processed optical fiber}


\author[1]{\fnm{Kailai} \sur{Liu}}
\equalcont{These authors contributed equally to this work.}

\author[1]{\fnm{Zhongyao} \sur{Luo}}
\equalcont{These authors contributed equally to this work.}

\author[1]{\fnm{Xiangpeng} \sur{Xiao}}
\equalcont{These authors contributed equally to this work.}

\author[1]{\fnm{Zhen} \sur{Wang}}

\author[1]{\fnm{Weiliang} \sur{Zhao}}

\author[1]{\fnm{Mingming} \sur{Zhang}}

\author*[1]{\fnm{Hao} \sur{Wu}}\email{wuhaoboom@hust.edu.cn}

\author*[1]{\fnm{Zhijun} \sur{Yan}}\email{yanzhijun@hust.edu.cn}

\author*[1,2]{\fnm{Ming} \sur{Tang}}\email{tangming@mail.hust.edu.cn}

\affil*[1]{\orgdiv{Wuhan National Laboratory for Optoelectronics, Next Generation Internet Access National Engineering Laboratory, and Hubei Optics Valley Laboratory, School of Optical and Electronic Information}, \orgname{Huazhong University of Science and Technology}, \orgaddress{\street{Luoyu Road No.1037}, \city{Wuhan}, \postcode{430074‌}, \country{China}}}

\affil[2]{ \orgname{Optical Fundamental Research Center}, \orgaddress{\street{Luoyu Road No.1037}, \city{Wuhan}, \postcode{430074‌}, \country{China}}}


\abstract{
As distributed sensing and computing systems exchange increasingly large volumes of data over optical fibers, the overhead of optoelectronic conversion, buffering, and scheduling in conventional transmit-then-compute architectures can approach or exceed the computation itself, making data movement a dominant bottleneck. Here, we propose FBG-CA, a passive fiber-Bragg-grating-based computing module fabricated using an automated through-coating inscription platform. During operation, spatially distributed weak FBGs reflect, delay, and accumulate different temporal samples to generate the convolution output, while the forward-propagating signal retains the original data stream and continues toward the remote receiver. Transmission and convolution therefore occur within the same optical propagation event, transforming the optical link from a passive transport channel into a distributed spatiotemporal computing medium for computation-in-flight. To improve robustness against finite reflection-level quantization and fabrication-induced deviations, we develop a hardware-aware training strategy that combines quantization aware training with sharpness aware minimization. Experimentally, the single-wavelength implementation achieves a convolution correlation coefficient of 0.987 and classification accuracies of 87.90\% and 79.07\% on the MNIST and MIT-BIH Arrhythmia datasets, respectively, with an estimated throughput of 288 GOPS over 1440 m of computing fiber. Despite an approximately 20\% forward-power reduction (about 0.97~dB) across the distributed FBG sections, the transmitted data retain high fidelity, with a mean structural similarity index measure of 0.9930 for MNIST images and a mean squared error of 0.005826 for MIT-BIH heartbeat waveforms. A four-channel wavelength division multiplexing implementation achieves 92.25\% accuracy on a four-class Fashion-MNIST task and an estimated throughput of 115.2~GOPS over 144 m of computing fiber. An architecture-level projection indicates an intrinsic throughput density of \mbox{1.6~POPS\,m\textsuperscript{\textminus 1}} for eight wavelength channels operating at 100~Gbaud per wavelength. These results establish FBG-CA as a passive and wavelength-scalable platform for computation-in-flight.
}




\maketitle

\section{Introduction}\label{sec1}

Modern sensing and computing infrastructures are increasingly organized as distributed systems, in which large volumes of data are continuously exchanged among sensors, storage servers, accelerator nodes, switches, and remote computing clusters. Optical fiber has become the dominant physical medium for these data flows, spanning long-haul backbone networks, metro and access links, data-center interconnects, and emerging chip, package, and rack-level optical links. In artificial-intelligence computing systems, optical communication is no longer a peripheral interface used only for long-distance transport. Instead, it is becoming an essential part of the computing fabric itself, because model parameters, activations, gradients, intermediate features, and routing decisions must be repeatedly exchanged across scale-up and scale-out architectures  \cite{miller2010optical,atabaki2018integrating,jouppi2023tpu,cheng2018recent,shekhar2024roadmapping}. 

Despite the rapid progress of electronic and photonic accelerators, data transmission and computation remain physically and temporally separated in conventional architectures. Data are first transported through optical links, then converted into electrical signals, buffered, scheduled, and processed by electronic accelerators. The processed results may subsequently be transmitted again to other nodes. This sequential transmit–then–compute paradigm introduces overhead from optical-to-electrical conversion, memory access, buffering, synchronization, and network scheduling. As the raw computing capability of accelerators continues to increase, data movement rather than arithmetic operation is becoming a dominant bottleneck in many distributed workloads. In large-scale deep-learning systems, cross-node communication and synchronization can substantially reduce accelerator utilization, and the communication time may become comparable to or even exceed the local computation time \cite{narayanan2021efficient,lao2021atp,hwang2023ark}. This limitation is particularly evident in distributed mixture-of-experts models, where repeated all-to-all token dispatch and collection can dominate the end-to-end training or inference latency  \cite{li2023accelerating}. 

These trends point toward a different architectural approach in which part of the computation is moved into the optical transmission path or brought as close to it as possible. If selected linear operations can be completed while the optical signal is propagating, then data transport and data processing no longer need to occur as strictly separated stages. Such a computation-in-flight architecture would treat the fiber not only as a communication channel but also as a distributed physical medium for signal processing. This idea is conceptually related to optical delay-line memories, where information is retained in continuously propagating optical signals \cite{sarrazin1990fiber, carmack2026fiber}. Recent fiber-based memory architectures have revisited this principle for streaming immutable model weights through recirculating optical links \cite{atmer2026fiber}. These developments suggest that optical information could be processed as it propagates, rather than merely stored in flight.

Photonic computing provides a natural physical basis for this transition because optical signals support high bandwidth, low propagation loss, wavelength parallelism, and analog linear operations. Existing photonic accelerators have demonstrated impressive capabilities, including coherent Mach–Zehnder interferometer (MZI) meshes for matrix–vector multiplication\cite{shen2017deep,xie2025complex}, wavelength-division-multiplexed (WDM) dispersion systems for high-speed convolution\cite{xu202111,bai2025tops}, and coherent photonic tensor processors for dynamic real-valued or complex-valued computation\cite{zhu2024silicon,zhu2024lightening,zhu2025coherent}. However, most of these platforms are implemented as dedicated computing modules rather than as parts of the transmission fiber. MZI meshes require precise phase control, calibration, and stabilization, which increases system complexity. The dispersion-based systems depend on dispersion engineering, wavelength spacing, and fiber length, which can restrict flexibility. Coherent photonic tensor processors usually require high-speed modulators, coherent receivers, digital-to-analog converters (DACs), analog-to-digital converters (ADCs), and large-scale radio-frequency (RF) control circuits. The active control and optoelectronic interfaces required by these architectures complicate their direct deployment as distributed computing elements within optical transmission links.

Here, we propose the Fiber-Bragg-grating-based convolution accelerator (FBG-CA), a passive all-fiber architecture that performs temporal convolution during optical transmission. Spatially distributed weak FBG sections encode convolution kernels through their longitudinal positions, Bragg wavelengths, and reflection strengths, allowing different temporal samples to be weakly reflected, delayed, and accumulated in the backward direction while the forward-propagating signal continues toward the remote receiver with limited impairment. To realize long, spatially distributed, and wavelength-selective optical kernels, we develop an automated precision inscription platform based on UV-transparent coated fiber (UV-TCF), in which synchronized fiber-length metrology, shutter gating, and phase-mask positioning translate digitally specified kernels into spatial, spectral, and multilevel reflection codes during continuous fiber translation. We further introduce a hardware-aware training strategy combining quantization-aware training (QAT) and sharpness-aware minimization (SAM) to improve robustness against finite reflection-level quantization and fabrication-induced deviations. Experimentally, the single-wavelength implementation achieves a convolution correlation coefficient of 0.987 and classification accuracies of 87.90\% and 79.07\% on the Modified National Institute of Standards and Technology (MNIST) dataset and the Massachusetts Institute of Technology--Beth Israel Hospital (MIT-BIH) Arrhythmia Database, respectively, with an estimated throughput of 288 giga-operations per second (GOPS) over 1440~m of computing fiber. The maximum reduction in forward optical power is approximately 20\%, corresponding to an insertion loss of 0.97~dB, while the transmitted signals retain high fidelity, with a mean structural similarity index (SSIM) of 0.9930 for MNIST images and a mean squared error (MSE) of 0.005826 for MIT-BIH heartbeat waveforms. A four-channel wavelength-division-multiplexed (WDM) implementation further achieves 92.25\% accuracy on a four-class Fashion-MNIST task and an estimated throughput of 115.2~GOPS over 144~m of computing fiber. At the architecture level, an eight-channel WDM system operating at 100~gigabaud (Gbaud) per wavelength is projected to reach an intrinsic throughput density of \mbox{1.6~POPS\,m\textsuperscript{\textminus 1}}. These results establish FBG-CA as a passive, wavelength-scalable, and transmission-compatible platform for computation-in-flight, providing a route toward embedding distributed linear processing directly into optical communication and sensing infrastructures.

\begin{figure}[h]
\centering
\includegraphics[width=1.0\textwidth]{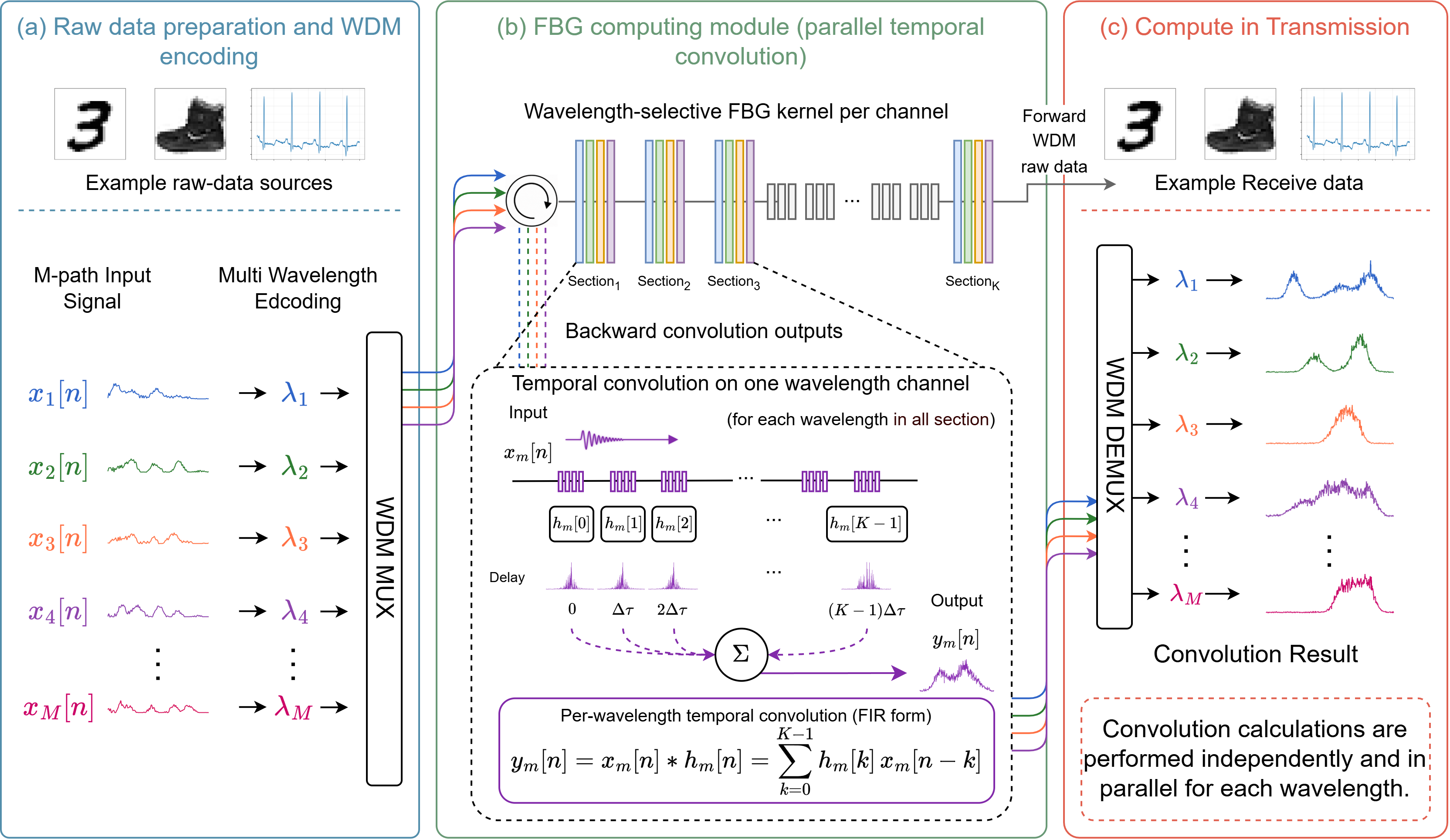}
\caption{Principle of FBG-CA for computation-in-flight.
(a) Temporal encoding of input data onto WDM optical carriers.
(b) Distributed wavelength-selective FBG sections performing parallel
temporal convolution during propagation.
(c) Integrated transmission and computation, where the forward signal
delivers the original data and the backward-reflected signal carries the
convolution results.}\label{mainfig}
\end{figure}

\section{Results}\label{sec2}
\subsection{Programmable FBG-CA weight mapping}\label{subsec1}
As shown in Fig. \ref{mainfig}, FBG-CA performs temporal convolution by mapping the discrete convolution kernel onto a series of wavelength-selective reflective structures inscribed along an optical fiber. A single FBG section can be fabricated by exposing a photosensitive fiber to a spatially periodic ultraviolet interference pattern or a phase mask, which induces a periodic refractive-index modulation in the fiber core. FBG-CA can be further extended to WDM parallel convolution by exploiting the wavelength selectivity of the phase-mask inscription process. Different positions on a chirped phase mask correspond to different local grating periods and hence different Bragg wavelengths. The signal model and its assumptions are given in Methods.

\begin{figure}[h]
\centering
\includegraphics[width=1.0\textwidth]{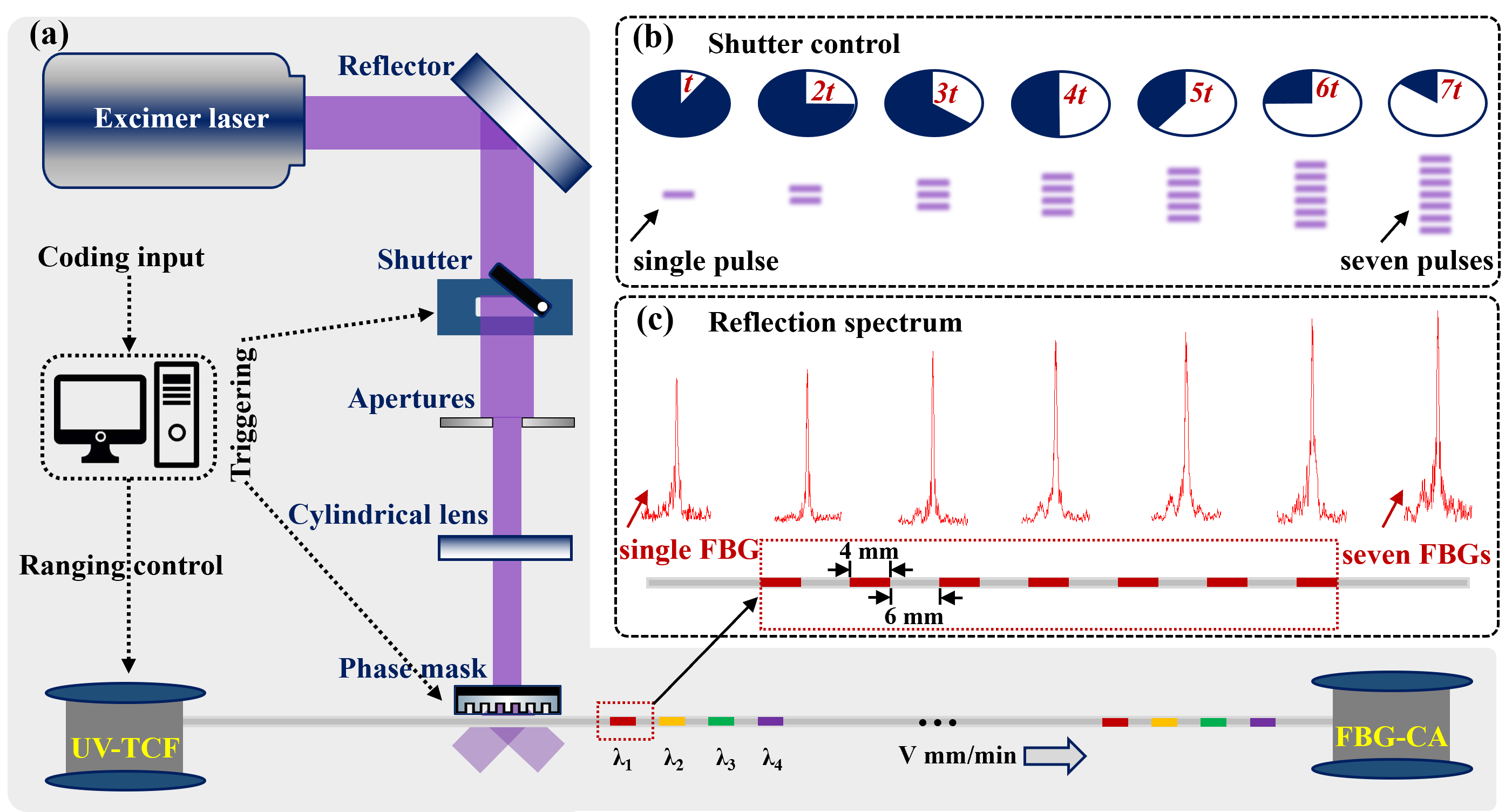}
\caption{Automated precision fabrication and multilevel weight encoding of FBG-CA. (a) Schematic of the automated through-coating FBG inscription platform. (b) Pulse-count encoding of seven nonzero reflection levels. (c) Measured superimposed reflection spectra for weight sections containing one to seven sub-FBGs. Each sub-FBG is 4 mm long and adjacent sub-FBGs are separated by a 10 mm pitch, corresponding to a 6 mm unexposed interval.}\label{FBG_Fab}
\end{figure}

Implementing this mapping over a long and continuous fiber requires substantially more than conventional point-by-point FBG inscription. Thousands of weak FBGs must be positioned at prescribed longitudinal coordinates while their wavelengths and reflectivities are independently programmed with sufficient repeatability. To meet these requirements, we developed an automated precision inscription platform based on UV-transparent coated fiber (UV-TCF), as illustrated in Fig. \ref{FBG_Fab}(a). The pulsed UV beam (248 nm) from an excimer laser is redirected by a UV-grade mirror, gated by a programmable shutter, and spatially shaped and focused onto the continuously translating UV-TCF through a phase mask. A fiber length metrology unit continuously registers the fiber position and synchronizes the shutter timing and phase-mask address with a digitally prescribed encoding file. The target positions, wavelength channels, and reflection levels can therefore be written automatically during continuous fiber translation.

The quantized optical weights are implemented through pulse count controlled sub-FBG aggregation, as shown in Fig. \ref{FBG_Fab}(b). At a fixed laser repetition rate and fiber translation speed, a shutter open duration of t admits one pulse and produces one sub-FBG, whereas durations from 2t to 7t admit two to seven pulses and generate the corresponding numbers of spatially separated sub-FBGs. Each sub-FBG is 4 mm long, and adjacent sub-FBGs are written at a pitch of 10 mm. Because the sub-FBGs have the same Bragg wavelength, their reflected contributions fall within the same temporal readout window and collectively form the effective reflectivity of one weight section. As experimentally shown in Fig. \ref{FBG_Fab}(c), increasing the sub-FBG number from one to seven produces seven clearly distinguishable and monotonically increasing reflection levels. Together with the unexposed state representing zero, this scheme realizes the eight level optical weight alphabet used in hardware-aware training.

The discrete FBG weight alphabet and residual fabrication deviations require a hardware-compatible mapping from trained network weights to the physical device. We therefore combined quantization-aware training (QAT) with sharpness-aware minimization (SAM) \cite{jacob2018quantization,foret2021sharpness} (Supplementary Fig. S1). The optimization procedure and measured-noise model are described in Methods. Under the measured perturbation model, QAT+SAM achieved the highest simulated accuracy (87.51\%) and the smallest degradation (2.2 percentage points).

\subsection{Serial implementation}\label{subsec2}

\begin{figure}[h]
\centering
\includegraphics[width=1\textwidth]{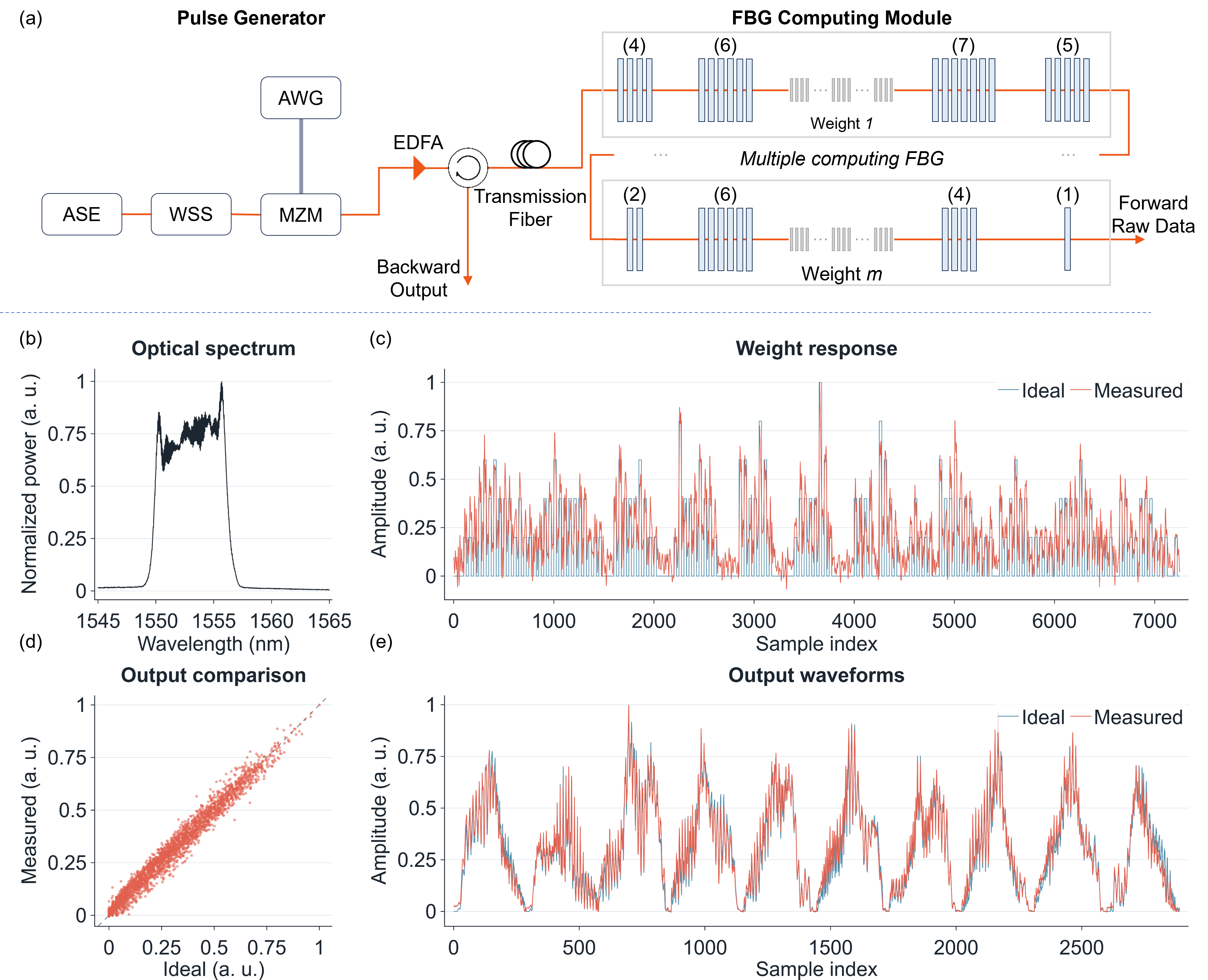}
\caption{Experimental setup and characterization of the serial FBG-CA architecture.
(a) Experimental setup of the serial FBG-CA system. ASE: Amplified Spontaneous Emission, WSS: Wavelength Selective Switch, MZM: Mach-Zehnder modulator, AWG: Arbitrary Waveform Generator, EDFA: Erbium Doped Fiber Amplifier.
(b) Measured optical spectrum of a representative FBG section.
(c) Comparison between the designed and measured FBG weight responses.
(d) Scatter plot of the measured convolution samples versus the reference values.
(e) Output waveform comparison between the ideal reference and the measured convolution result using the first MNIST test sample.
}\label{mnistsetup}
\end{figure}

Fig. \ref{mnistsetup}(a) shows the experimental implementation of the serial FBG-CA system. Spatially distributed FBG weight sections are separated by 1 m, producing an approximately 10-ns round-trip delay that matches the temporal interval of the input sequence. The backward-reflected replicas from these sections are weighted, delayed, and accumulated to generate the convolution output, while the weakly perturbed forward-propagating signal continues toward the remote receiver. The serial FBG-CA therefore performs temporal convolution and data transmission simultaneously within the same optical propagation process, enabling passive computation-in-flight without interrupting the main transmission path.

Fig. \ref{mnistsetup}(b) shows the measured optical spectrum of a representative FBG section, exhibiting a bandwidth larger than 5 nm. This broad spectral response relaxes the requirement on optical source wavelength stability and enables the system to tolerate moderate wavelength drift. Fig. \ref{mnistsetup}(c) compares the designed and measured weight responses encoded in the serial FBG computing fiber. The measured reflection amplitudes generally follow the target weight distribution, confirming that the prescribed weights can be accurately mapped onto the passive FBG reflection profile. To quantitatively evaluate the convolution accuracy, Fig. \ref{mnistsetup}(d) plots the measured convolution samples against the corresponding reference values. The data points are tightly distributed along the diagonal, with a correlation coefficient of 0.987, demonstrating the high fidelity of the serial FBG computing fiber in reproducing the desired convolution operation. Fig. \ref{mnistsetup}(e) further shows the convolution result obtained using the first test sample from the MNIST dataset. The experimentally measured output waveform closely follows the ideal reference waveform calculated from the same input and the designed FBG weights, indicating that the serial FBG-CA module can correctly perform passive temporal convolution on real image-derived pulse sequences.

\subsection{Optical inference and classification readout}
\label{subsec:readout}

\begin{figure}[h]
\centering
\includegraphics[width=1.0\textwidth]{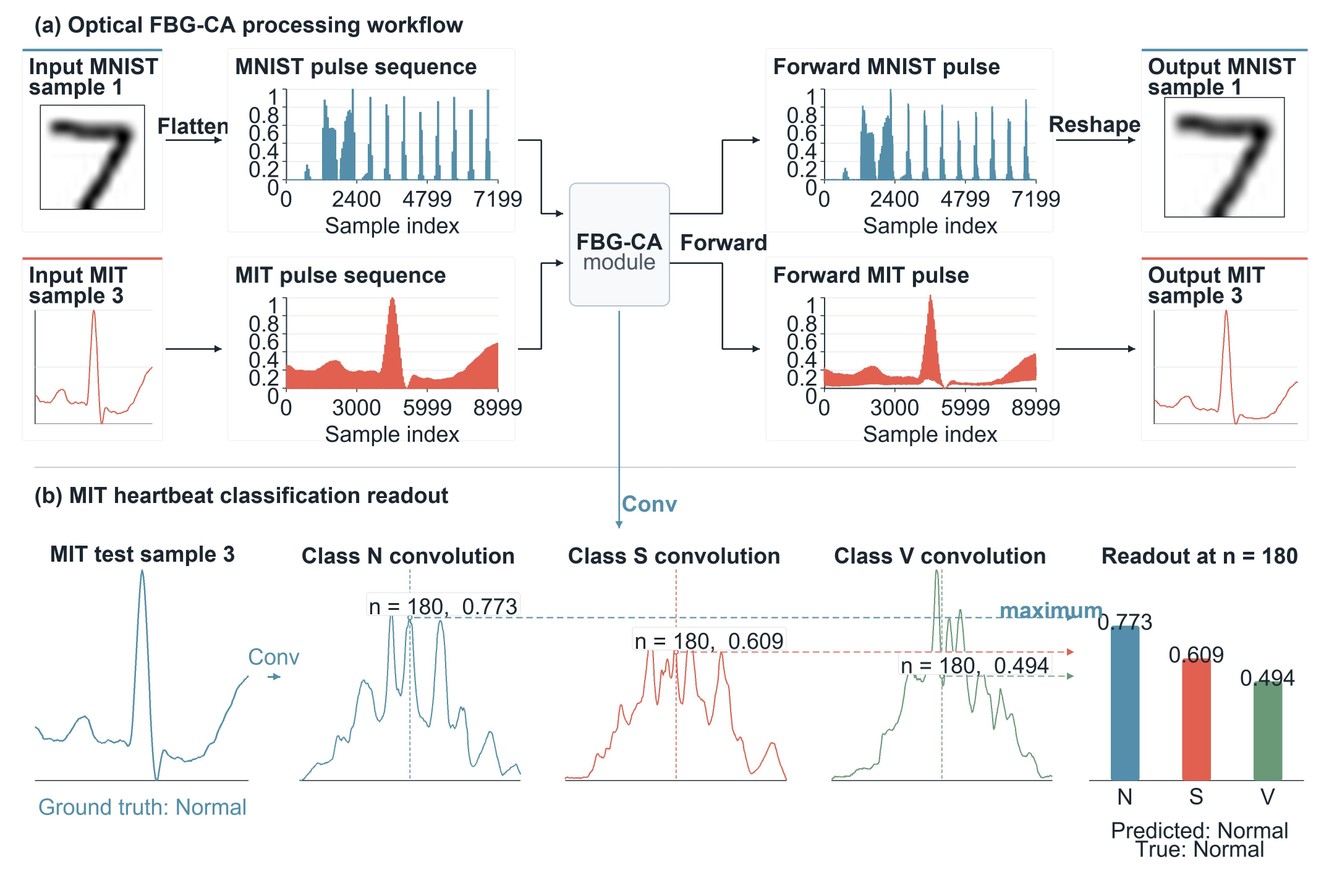}
\caption{Optical processing and classification readout of the serial FBG-CA system.
(a) Optical processing workflow for the MNIST and MIT-BIH Arrhythmia datasets, showing that the forward-propagating pulse sequences preserve the original input data after transmission through the FBG-CA module.
(b) Experimental classification of an MIT-BIH heartbeat sample. The input waveform is convolved with three class-specific FBG weight profiles corresponding to the Normal, Supraventricular, and Ventricular classes. The maximum measured response at the designated readout position determines the predicted class.}
\label{mainresult}
\end{figure}

The serial FBG-CA module was evaluated on MNIST and the MIT-BIH Arrhythmia Database \cite{goldberger2000physiobank,moody2001impact}. MNIST is a widely used benchmark dataset containing 70,000 grayscale images of handwritten digits from 0 to 9, while the MIT-BIH Arrhythmia dataset is a standard electrocardiogram (ECG) database for arrhythmia analysis. As illustrated in Fig. \ref{mainresult}(a), samples from both datasets were converted into one-dimensional temporal pulse sequences and modulated onto the optical carrier using a Mach-Zehnder modulator (MZM). The forward-propagating signal preserves the original temporal pulse sequence and can be reshaped to recover the input MNIST image or MIT heartbeat waveform. This result demonstrates that the FBG-CA module remains transparent to the main transmission path while simultaneously generating class-specific convolution results in the backward-reflected branch. In the single-wavelength MNIST implementation, the forward signal traverses approximately $2\times10^3$ weak FBG elements. With an individual reflectivity of approximately $10^{-4}$, the cumulative extraction corresponds to an approximately 20\% reduction in forward optical power, or an insertion loss of about 0.97~dB. Despite this modest loss, the reconstructed MNIST images achieve a mean SSIM of 0.9930 over ten representative test samples. For the corresponding ten MIT heartbeat waveforms, the MSE between the normalized input and recovered output sequences is 0.005826. These results confirm that the forward branch retains high-fidelity image structure and heartbeat waveform transmission under the distributed FBG extraction loss.

Fig. \ref{mainresult}(b) illustrates the complete optical classification process using the third MIT-BIH test sample, whose ground-truth label is N. The input heartbeat waveform is independently convolved with three class-specific weight sequences corresponding to the Normal, Supraventricular, and Ventricular heartbeat classes. These three convolution traces are experimentally measured outputs generated by the fabricated FBG computing fiber, rather than numerically reconstructed results. For each class branch, the reflected optical waveform represents the temporal convolution between the input heartbeat sequence and the corresponding FBG-encoded weight profile.

The class scores are obtained by sampling the three measured convolution outputs at the designated readout position, $n=180$. At this position, the measured responses for the Normal, Supraventricular, and Ventricular classes are 0.773, 0.609, and 0.494, respectively. The N-class branch therefore produces the maximum response, and the sample is classified as N, in agreement with its ground-truth label. This maximum-response readout provides a direct winner-take-all classification mechanism, in which the output class is determined by comparing the class-specific convolution scores generated by the physical FBG kernels. The result demonstrates that the fabricated serial FBG-CA can complete the full inference process, including optical convolution, class-score generation, and final classification readout.

The serial FBG-CA system achieved experimental classification accuracies of 87.90\% on ten-class MNIST and 79.07\% on three-class MIT-BIH, compared with 89.75\% and 79.32\%, respectively, for the corresponding GPU-based numerical models. The class-response maps and confusion matrices are provided in Supplementary Fig. S3.

\subsection{WDM Extension}\label{subsec3}

\begin{figure}[h]
\centering
\includegraphics[width=1\textwidth]{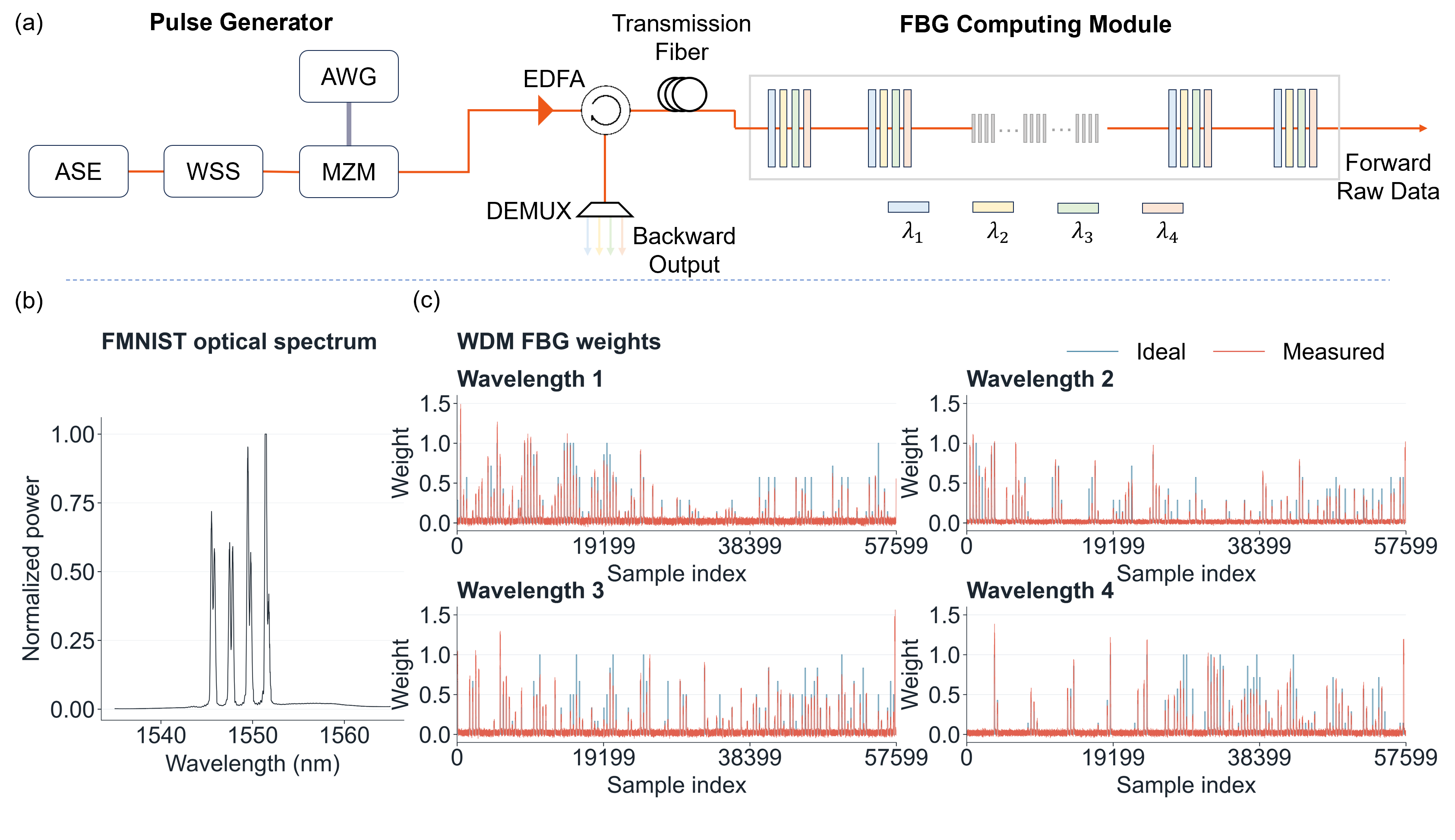}
\caption{WDM implementation of FBG-CA. (a) Experimental setup of the WDM FBG-CA architecture. (b) Measured optical spectrum of the WDM input signal. (c) Comparison between the designed and measured FBG weights for four wavelength channels.}
\label{WDMsetup}
\end{figure}

Building on the wavelength-selective kernel encoding described above, Fig. \ref{WDMsetup}(a) shows the WDM FBG-CA implementation. Independent wavelength channels interact with their corresponding FBG kernels and produce parallel convolution outputs while the forward WDM signal continues through the transmission link. WDM operation also increases the computational density of FBG-CA by co-locating multiple kernels within the same longitudinal fiber region. With $N$ wavelength channels, $N$ convolution kernels can operate in parallel, ideally reducing the required fiber length for a fixed computational workload to $1/N$ of that of the single-wavelength implementation. Equivalently, for a fixed fiber length, WDM provides an ideal $N$-fold increase in the number of convolution operations and aggregate throughput. Because wavelength multiplexing is already widely used in optical interconnects, this scaling approach can leverage existing sources, multiplexers, demultiplexers, and standardized wavelength grids to incorporate parallel computation-in-flight into optical transmission systems.

Figs. \ref{WDMsetup}(b)-(c) characterize the WDM FBG-CA module and verify its wavelength-parallel computing capability. The system was evaluated using Fashion-MNIST (FMNIST) \cite{xiao2017fashion}, a ten-class grayscale image dataset of fashion products with the same image size and data format as MNIST. Four representative classes (0, 1, 2 and 5) were retained and remapped to four output classes. Fig. \ref{WDMsetup}(b) shows the measured optical spectrum of the resulting WDM input signal, where multiple subcarriers are clearly resolved within the 1550-nm band. Fig. \ref{WDMsetup}(c) compares the designed and measured FBG weight responses for the four wavelength channels. The measured reflection peaks generally follow the target weight distributions in all channels, confirming that the chirped-FBG fabrication process can assign different reflection profiles to different wavelengths within the same computing fiber. The measured convolution outputs exhibit strong correlations with the numerical references, with correlation coefficients of 0.968, 0.926, 0.917, and 0.954 for wavelength channels 1--4, respectively. These results demonstrate that WDM FBG-CA can perform multiple convolution operations in parallel across different wavelength channels, validating its capability for high-density, passive, and transmission-compatible optical computing.

\begin{figure}[h]
\centering
\includegraphics[width=1\textwidth]{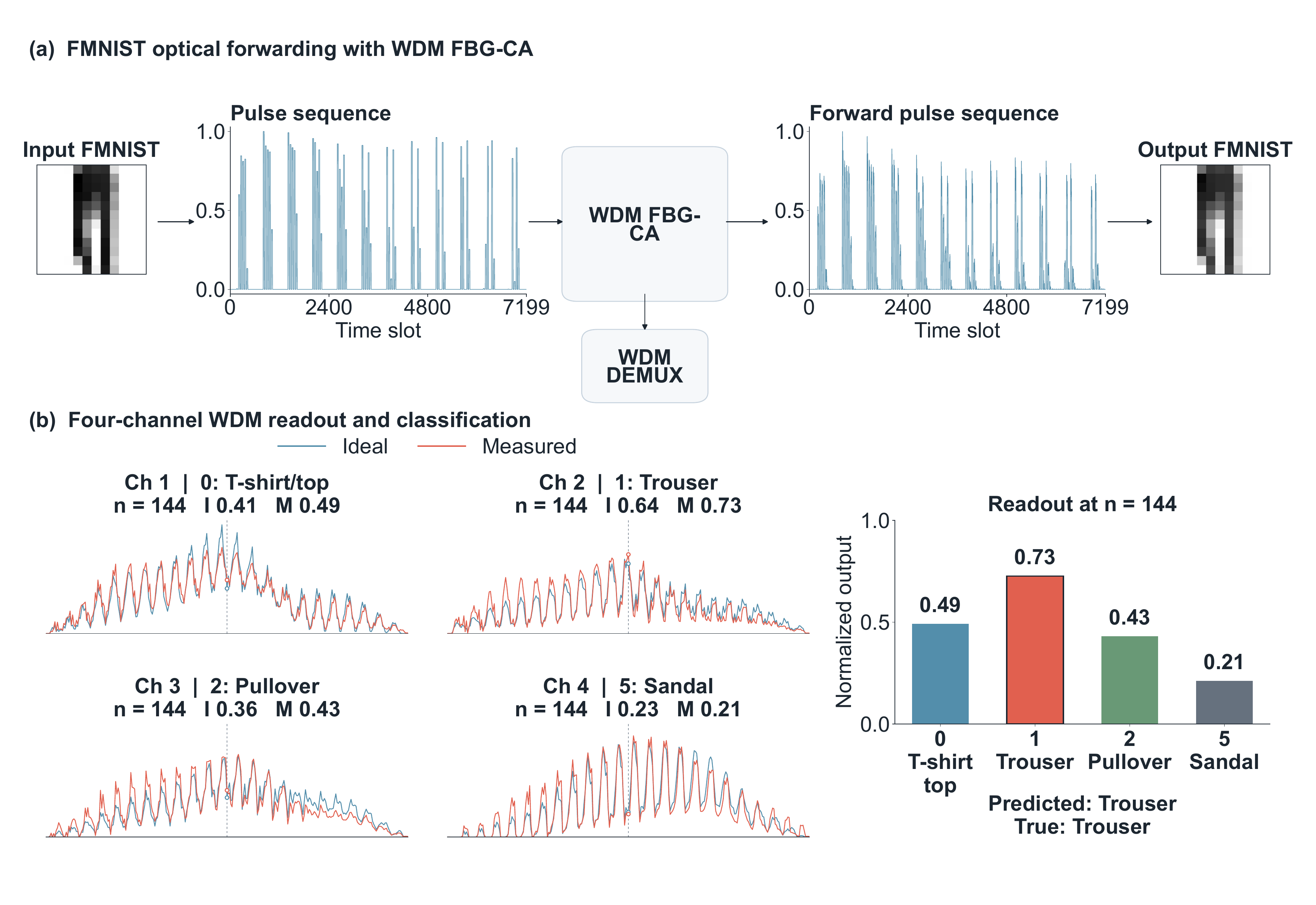}
\caption{Multi-wavelength convolution and FMNIST classification using WDM FBG-CA.
(a) Optical forwarding workflow of the FMNIST pulse sequence through the WDM FBG-CA module.
(b) Demultiplexed WDM convolution outputs and the corresponding FMNIST classification result.}
\label{wdmfmnist}
\end{figure}

The complete optical forwarding and classification workflow is shown in Fig. \ref{wdmfmnist}. As illustrated in Fig. \ref{wdmfmnist}(a), the preprocessed 144-sample sequence is jointly modulated onto multiple WDM carriers. After propagating through the WDM FBG-CA module, the forward signal preserves the input sequence and can be reshaped to recover the original image at the output. Meanwhile, the backward-reflected branch carries the wavelength-parallel convolution results. This simultaneous preservation of the forward data stream and extraction of the reflected computing outputs confirms that the WDM FBG-CA module remains transparent to the normal transmission path while performing convolution during propagation. For the FMNIST implementation, each wavelength channel traverses approximately $10^3$ weak FBG elements in the forward path. The reconstructed images achieve a mean SSIM of 0.9774 over ten representative samples, confirming that the forward transmission preserves the image structure with only minor distortion.

Fig. \ref{wdmfmnist}(b) shows the demultiplexed convolution outputs from the four WDM channels. Each wavelength channel is reflected by its corresponding wavelength-selective FBG weights and independently performs temporal convolution with the input pulse sequence. After WDM demultiplexing, the measured output waveform of each channel is compared with the numerically calculated reference result. The measured traces generally follow the reference curves, indicating that the WDM FBG-CA module can execute multiple convolution operations in parallel across different wavelength channels. The demultiplexed convolution scores were subsequently used for four-class FMNIST classification, yielding an experimental accuracy of 92.25\%. For comparison, the corresponding GPU-based numerical computation achieved an accuracy of 94.82\%. The moderate accuracy degradation relative to the digital reference, as well as the more noticeable degradation compared with the single-wavelength implementation, can be mainly attributed to the increased fabrication complexity of the WDM FBG-CA device. In the WDM implementation, each FBG section must simultaneously control both the reflection wavelength and reflection strength of multiple sub-FBGs, making the system more sensitive to wavelength-position errors, reflectivity deviations, spectral overlap, inter-channel crosstalk, and wavelength-dependent loss. Despite these nonidealities, the WDM FBG-CA module maintains high classification accuracy and demonstrates the feasibility of high-density, wavelength-parallel, passive optical convolution.

\section{Discussion}\label{sec:discussion}

The key contribution of FBG-CA is the integration of optical transmission and computation within the same fiber medium. Unlike conventional architectures, in which optical data must first be converted, buffered, and transferred to electronic processors, FBG-CA performs weighted delay-and-accumulation operations directly during propagation. The forward-propagating signal delivers the original data stream to the remote receiver, while the backward-reflected branch simultaneously carries the convolution result, enabling the transmission medium itself to participate in data processing through computation-in-flight. An enabling element of this architecture is the automated through-coating fabrication process. Rather than relying on the fabrication and assembly of discrete optical components, the platform directly compiles a digitally specified kernel into distributed refractive-index structures along a continuous fiber. Longitudinal position, Bragg wavelength, and pulse-count-controlled reflectivity thereby constitute three jointly programmable physical degrees of freedom for optical computation. At a conceptual level, FBG-CA extends the fundamental role of optical fiber beyond data transmission by establishing it as a computing medium for in-flight information processing.

The computational capacity of FBG-CA is jointly determined by the longitudinal density of the FBG computing units and the number of wavelength channels. Let $X$ denote the length of the computing fiber, $\rho$ the number of FBG computing units per unit length, and $c_f$ the optical group velocity in the fiber. The total number of FBG computing units is therefore $N_{\mathrm{FBG}}=X\rho$. Because the round-trip delay between adjacent units defines the temporal sampling interval, the symbol period and symbol rate are given by

\begin{equation}
T_s=\frac{2}{\rho c_f},
\qquad
f_s=\frac{1}{T_s}=\frac{\rho c_f}{2}.
\end{equation}

During steady-state streaming operation, each FBG computing unit performs one multiplication and one accumulation for every input symbol, corresponding to two operations. Since all $X\rho$ units operate in parallel at a symbol rate of $\rho c_f/2$, the intrinsic single-wavelength throughput is

\begin{equation}
R_{\mathrm{1\lambda}}
=
2N_{\mathrm{FBG}}f_s
=
2(X\rho)\left(\frac{\rho c_f}{2}\right)
=
X\rho^2c_f.
\end{equation}

The intrinsic optical latency is determined by the round-trip propagation time to the farthest FBG section,

\begin{equation}
\tau_{\mathrm{opt}}=\frac{2X}{c_f}.
\end{equation}

Increasing the fiber length accommodates more computing units and therefore increases the steady-state throughput, but it also increases the propagation latency. By contrast, increasing the FBG density not only places more computing units within a given fiber length but also raises the symbol rate, resulting in a quadratic throughput scaling with density. When $M$ WDM channels are employed, each wavelength independently performs convolution using its own FBG-encoded kernel, and the ideal aggregate throughput becomes

\begin{equation}
R_{\mathrm{WDM}}
=
MR_{\mathrm{1\lambda}}
=
MX\rho^2c_f.
\end{equation}

Accordingly, for a fixed fiber length and group velocity, the intrinsic throughput scales quadratically with the FBG density, or equivalently with the symbol rate, as $R\propto\rho^2\propto f_s^2$, while WDM provides an additional linear scaling factor of $M$. This multiplicative scaling across the spatial, temporal, and wavelength dimensions offers substantial computational headroom for future high-density FBG-CA systems. The above estimates assume continuous streaming and full pipeline utilization and do not include additional latency or bandwidth constraints arising from modulation, detection, data conversion, and electronic readout.

The fabricated computing fiber operates at a unit density of $\rho=1~\mathrm{m}^{-1}$ and a symbol rate of $f_s=\rho c_f/2=100$ megabaud (MBd), corresponding to an intrinsic throughput density of $\rho^2 c_f=0.2~\mathrm{GOPS\,m^{-1}}$ per wavelength channel. Under the experimentally demonstrated conditions, the single-wavelength MNIST implementation comprises $144\times10$ computing units distributed over 1440 m of fiber, corresponding to an estimated aggregate throughput of 288 GOPS and an intrinsic round-trip latency of $14.4~\mu\mathrm{s}$. The four-channel WDM Fashion-MNIST implementation encodes $144\times4$ computing units within 144 m of fiber, i.e., $0.8~\mathrm{GOPS\,m^{-1}}$, yielding an estimated aggregate throughput of 115.2 GOPS and a latency of $1.44~\mu\mathrm{s}$. To illustrate the scaling potential, a symbol rate of 100 gigabaud (Gbaud) per wavelength corresponds to an adjacent-unit spacing of $\Delta z=c_f/(2f_s)=1~\mathrm{mm}$ and an FBG density of $\rho=1000~\mathrm{m}^{-1}$. With eight WDM channels, the resulting computational throughput density reaches $R_{\mathrm{WDM}}/X=M\rho^2c_f=1.6\times10^{15}~\mathrm{operations\,s^{-1}\,m^{-1}}$, equivalent to a throughput density of 1.6 POPS~m$^{-1}$. This value represents an ideal scaling estimate, whereas the practically achievable throughput density will also be constrained by the FBG fabrication pitch and precision, transmitter and receiver bandwidths, accumulated optical loss, available dynamic range, chromatic dispersion, and inter-channel crosstalk.

The transmission-compatible topology of FBG-CA is applicable to scenarios
in which optical links naturally coexist with fixed linear processing.
Potential examples include confidence estimation and routing in
AI-cluster interconnects \cite{teerapittayanon2016branchynet,
shazeer2017outrageously,fedus2022switch}, analog signal processing in
radio-over-fiber links \cite{beas2013millimeter}, and preprocessing of
fiber-sensing signals \cite{gao2025optical,gao2024breaking}. In these
cases, operations such as correlation, matched filtering, feature
extraction, and anomaly detection can be performed during propagation
rather than introduced as separate processing stages
\cite{zhong2021ioi,sludds2022delocalized}.

The encoded weights are nonvolatile refractive-index structures and
therefore require neither refresh energy nor static tuning power. The
present architecture is most suitable for fixed or slowly reconfigurable
kernels, although thermal or strain tuning may provide post-fabrication
calibration and limited adjustment of the Bragg response
\cite{kersey1997fiber,campanella2018fibre}. Signed weights could be
supported through differential encoding, balanced detection,
dual-wavelength channels, or offset-based representations
\cite{li2019class,tait2022quantifying}. Further scaling will require
higher-density grating fabrication, accurate spectral calibration, and
improved control of accumulated loss and wavelength-channel crosstalk.

In summary, we demonstrated FBG-CA, a passive all-fiber computing medium
that performs temporal convolution during optical transmission. The
single-wavelength implementation achieved a convolution correlation
coefficient of 0.987 and classification accuracies of 87.90\% and
79.07\% on the MNIST and MIT-BIH Arrhythmia datasets, respectively,
while the four-channel WDM implementation achieved 92.25\% accuracy on a
four-class Fashion-MNIST task. These results demonstrate that distributed
reflectivity, longitudinal delay, and wavelength selectivity can be
jointly exploited as physical computational resources, providing a route
toward embedding selected processing functions directly into optical
transmission links.

\section{Methods}

\subsection{FBG-CA signal model}

For a single wavelength channel, the input intensity sequence is denoted by $x[n]$ and the designed reflectivity of the $k$-th FBG section by $h[k]$. Neglecting higher-order reflections and treating propagation loss as either pre-compensated or included in $h[k]$, the backward reflected intensity collected by the circulator is
\begin{equation}
y[n] = \sum_{k=0}^{K-1} h[k] \cdot x[n-k].
\end{equation}
This is a finite-impulse-response convolution between the input sequence and the FBG-defined kernel. The system uses a broadband optical source, so reflected components from different sections are added primarily in optical intensity rather than through phase-sensitive coherent interference. The convolution output is read from the optical power at the photodetector, whereas the non-reflected component continues in the forward transmission path. The Bragg wavelength is $\lambda_{\mathrm{B}} = 2 n_{\mathrm{eff}}\Lambda$, where $n_{\mathrm{eff}}$ is the effective modal index and $\Lambda$ is the grating period. Adjacent FBG sections separated by $\Delta z$ generate a round-trip delay $\Delta \tau = 2n_g\Delta z/c$, the input sampling period was matched to this delay.

For the $m$-th WDM channel at wavelength $\lambda_m$, a separate reflectivity sequence $h_m[k]$ is encoded, yielding $y_m[n]=\sum_{k=0}^{K-1}h_m[k]x_m[n-k]$. The wavelength-selective FBG sections therefore implement co-located, independent convolution kernels for the WDM channels.

\subsection{Hardware-aware training and fabrication-noise model}

The neural-network weights were mapped to the available reflection levels using quantization-aware training (QAT). A full-precision weight $w$ was mapped to a hardware-compatible value as
\begin{equation}
\tilde{w}=Q(w)=s\,\mathrm{clip}\left(\mathrm{round}\left(\frac{w}{s}\right),q_{\min},q_{\max}\right),
\end{equation}
where $s$ is the quantization step and $q_{\min}$ and $q_{\max}$ define the available reflection levels. The task loss was evaluated using quantized weights, $\mathcal{L}_{\mathrm{QAT}}(\mathbf{w})=\mathcal{L}(Q(\mathbf{w}))$. The non-differentiable rounding operation was handled with a straight-through estimator during backpropagation.

Sharpness-aware minimization (SAM) was combined with QAT to improve tolerance to fabrication and measurement perturbations. SAM minimizes the worst-case loss within an $\ell_2$-bounded neighbourhood of the model parameters,
\begin{equation}
\mathcal{L}_{\mathrm{SAM}}(\boldsymbol{\theta})=\max_{\|\boldsymbol{\epsilon}\|_2\leq\rho}\mathcal{L}(\boldsymbol{\theta}+\boldsymbol{\epsilon}),
\end{equation}
with the adversarial perturbation approximated at each iteration by
\begin{equation}
\hat{\boldsymbol{\epsilon}}=\rho\frac{\nabla_{\boldsymbol{\theta}}\mathcal{L}(\boldsymbol{\theta})}{\left\|\nabla_{\boldsymbol{\theta}}\mathcal{L}(\boldsymbol{\theta})\right\|_2}.
\end{equation}
The QAT+SAM strategy was selected for the fabricated device because it provided the most robust performance under the measured FBG-weight deviations. The measured deviations and the comparative robustness ablation are provided in Supplementary Fig. S2.

\subsection{Datasets and preprocessing}

For MNIST \cite{lecun1998gradient}, all ten digit classes were evaluated. Original $28\times28$ grayscale images were centre-cropped to $20\times20$, resized to $12\times12$, and flattened to 144-element input vectors. The single-layer $144\times10$ model reduced the number of optical weights from $784\times10=7840$ to $144\times10=1440$.

For the MIT-BIH Arrhythmia Database \cite{goldberger2000physiobank,moody2001impact}, the first ECG channel of each record was sampled at 360 Hz. A beat-centred segment containing 90 samples before and 90 samples after each valid annotation was extracted to form a 180-point sequence (approximately 0.5 s), then independently min--max normalized to $[0,1]$ without further downsampling or binarization. The Normal, Supraventricular and Ventricular beat classes were selected for the three-kernel physical implementation.

For the WDM experiment, FMNIST \cite{xiao2017fashion} inputs were preprocessed in the same manner as MNIST. T-shirt/top, Trouser, Pullover and Sandal (original labels 0, 1, 2 and 5) were retained and remapped to four output classes.

\subsection{Optical experiments and input encoding}

For serial FBG-CA experiments, a broadband amplified spontaneous emission (ASE) source was spectrally filtered by a wavelength-selective switch (WSS) to select a 1-nm-wide slice centred in the 1550-nm telecommunications band. The ASE-WSS configuration was adopted mainly for experimental convenience. In practical implementations, it can be replaced by compact broadband or low-coherence optical sources, such as superluminescent diodes (SLDs) \cite{lu2024low}, semiconductor optical amplifier (SOA)-based ASE sources \cite{yeh2012ultra}, or highly multimode/random semiconductor lasers \cite{redding2015low,lu2024low}, provided that their optical spectra cover the reflection band of the FBG computing module. Input sequences were converted into one-dimensional temporal pulse sequences and modulated using a Mach--Zehnder modulator (MZM) driven by an arbitrary waveform generator (AWG). A lookup-table predistortion was applied to compensate the nonlinear electro-optic transfer function of the MZM \cite{yang2023digital}. The modulated signal was amplified by an erbium-doped fiber amplifier (EDFA) and injected into the FBG-CA module through an optical circulator. The backward-reflected branch was detected as the computing output and the forward branch was used to assess transmission fidelity.

For WDM experiments, the ASE source was shaped by the WSS to generate multiple wavelength channels, which were jointly modulated by the same input sequence. Each wavelength interacted with its assigned FBG kernel. The backward-reflected WDM outputs were separated by a wavelength demultiplexer and detected individually, while the forward WDM signal continued through the link.

\bmhead{Acknowledgements}

This research is supported by the National Key Research and Development Program of China(2023YFB2906300); National Natural Science Foundation of China (62225110 and 62205114); JD project of Hubei province (2023BAA013); Hubei Provincial Natural Science Foundation of China (2025AFB008).

\bmhead{Author contributions}
M. T. and H. W. proposed the idea and the conceptual design. K. L. and Z. L. built the experiments, optimized the system, and performed the measurements. M. Z. assisted with the experiments. X. X., W. Z., Z. W., and Z. Y. designed and manufactured the FBG. K. L. and H. W. wrote the paper. All authors revised the paper.

\bmhead{Data availability}

The datasets, source code, and data-processing scripts
supporting this study are available in the GitHub repository:
\href{https://github.com/HUST-IOF/FBG-CA}{https://github.com/HUST-IOF/FBG-CA}.

\bmhead{Conflict of interest}
The authors declare no competing interests.









\bibliography{sn-bibliography}

\end{document}